# *Ab initio* modelling of UN grain boundary interfaces

Eugene A Kotomin[1], Yuri F Zhukovkii[1], Dmitry Bocharov[*,2,3], Denis Gryaznov[1]

[1]Institute of Solid State Physics, University of Latvia, Kengaraga Str. 8, LV-1063, Riga, Latvia.
[2]Faculty of Computing, University of Latvia, Raina blvd. 19, LV-1586, Riga, Latvia.
[3]Faculty of Physics and Mathematics, University of Latvia, Zellu Str. 8, LV-1002, Riga, Latvia.

E-mail: bocharov@latnet.lv

**Abstract.** The uranium mononitride (UN) is a material considered as a promising candidate for Generation-IV nuclear reactor fuels. Unfortunately, oxygen in air affects UN fuel performance and stability. Therefore, it is necessary to understand the mechanism of oxygen adsorption and further UN oxidation in the bulk and at surface. Recently, we performed a detailed study on oxygen interaction with UN surface using density functional theory (DFT) calculations. We were able to identify an atomistic mechanism of UN surface oxidation consisting of several important steps, starting with the oxygen molecule dissociation and finishing with oxygen atom incorporation into vacancies on the surface. However, in reality most of processes occur at the interfaces and on UN grain boundaries. In this study, we present the results of first DFT calculations on O behaviour inside UN grain boundaries performed using GGA exchange-correlation functional PW91 as implemented into the *VASP* computer code. We consider a simple interface (310)[001](36.8°) tilt grain boundary. The N vacancy formation energies and energies of O incorporation into pre-existing vacancies in the grain boundaries as well as O solution energies were compared with those obtained for the UN (001) and (110) surfaces.

## 1 Introduction

Uranium mononitride (UN) is a compound with metallic properties (metallic colour and low electrical resistivity) [1], possessing rock-salt *fcc* structure over a wide temperature range. UN is considered nowadays by the Generation IV International Forum of nuclear reactors [2] as one of the most promising future nuclear fuels alternative to currently used $UO_2$. However, UN reveals unwanted oxidation in air which could affect the fuel fabrication process and fuel performance [3]. Due to a considerable amount of chemically aggressive oxygen impurities in any UN samples, it is necessary to understand the atomistic mechanism of O adsorption and further oxidation process.

The detailed study of single-crystalline UN surface including their interaction with oxygen have been performed by us recently [4] using the DFT method as implemented in the VASP computer code [5]. We performed such the calculations for a pure UN bulk and surface, nitride and uranium vacancies on the surface, atomic and molecular adsorption as well as oxygen diffusion on the UN (001) surface (we studied mainly the (001) surface since according to the traditional Tasker analysis [6] it has the lowest surface energy).

---
[*] To whom any correspondence should be addressed.

The results obtained for interaction of O atoms and oxygen molecules with the UN (001) surface [4] demonstrate strong chemisorption, typical for metallic adsorbents. The possibility for spontaneous dissociation of the adsorbed $O_2$ molecules upon the perfect UN surface, analogous to the $O_2$ dissociation on metallic surfaces, has been demonstrated. After molecular dissociation, O adatom ($O_{ads}$) forms a strong chemical bond with U atom in the outermost surface plane ($U_{surf}$) which can be considered as one-center surface complex. High mobility of $O_{ads}$ atoms along the surface due to relatively low migration barriers (<0.5 eV) has been found. The possibility of low-barrier (~0.5-1 eV) for $O_{ads}$ incorporation (from the nearest adsorption site atop $U_{surf}$) into pre-existing N vacancy has been also shown as well as energetic stability of UN surface containing incorporated O atoms [4]. To increase validity of results, we have additionally performed the (110) surface calculations. We have compared the energies of N vacancy formation, oxygen atom adsorption upon U or N surface atom as well O atom incorporation into the N vacancy evaluated for both the UN(001) and (110) surfaces [4].

Nevertheless, many questions on UN oxidation remain open. Taking into account the fact that polycrystalline UN fuel powder contains nanoparticles with differently oriented crystallographic facets and a wide variety of interfaces between the grains [7], we consider here a role of grain boundaries (GB) in UN oxidation. This makes theoretical simulation of grain boundaries between different UN facets and their interaction with defects and O impurities very important for a realistic description of actinides. It is also important to study similarities and differences in oxygen incorporation processes energetic on both single surface and grain boundary.

In this paper, we consider a simple grain boundary interface (310)[001](36.8°) model and present first results for the energetics of N vacancy formation in the grain boundaries as well as oxygen incorporation into pre-existing N vacancies.

## 2 Computational details

The results were obtained using the plane wave calculations with *VASP* computer code, employing a plane-wave basis set combined with *PAW* pseudopotentials for U, N and O atoms (containing 14, 5 and 6 valence electrons, respectively) [5].

In this study, we have generated *k*-points using Monkhorst-Pack's technique [8] whereas the electron populations were determined following the method of Methfessel and Paxton [9] as implemented in the *VASP* code. For each series of calculations, we have found the optimal k-point mesh that provides convergence of the results: the 4×4×4 mesh for grain boundary calculations (the 4×4×1 and 8×8×1 meshes was used for previous surface calculations).

We have performed spin-relaxed calculations of ferromagnetic UN. The smearing parameter of 0.2 eV has been found to be optimal for reasonable convergence suggesting the electronic entropy contribution of the order of 10 meV. The optimal cut-off energy has been found to be equal to 520 eV. In this study, we present results of first DFT calculations on oxygen behaviour between UN grain boundaries.

## 3 Model

We consider a model of (310)[001](36.8°) tilt grain boundary used previously for describing of electron-trapping in rock-salt compounds like MgO, NaCl and LiF [10] a slab model for surface modeling. The periodic supercell with 15.40 Å × 4.87 Å × 34.13 Å linear dimensions contains 160 atoms (or 159 for N vacancy calculations). We have considered three sites for the N vacancies in positions of N atoms enumerated as 1 to 3 in Fig. 1. The lattice constant of UN slabs is fixed at 4.868 Å, taken from the lattice relaxation of UN bulk [11]. In all calculations we performed a complete structure optimization inside the supercell of fixed linear dimensions using criterion of the total energy minimization.

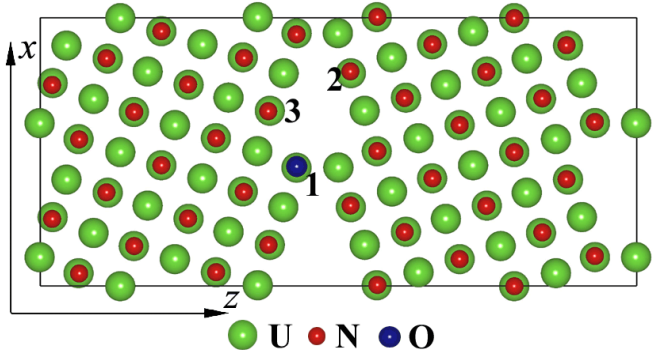

*Figure 1.* Cross-section of the (310)[001](36.8°) tilt grain boundary supercell (15.40 Å × 4.87 Å × 34.13 Å) for UN slab model with oxygen incorporated into one of possible positions (here N atom in position 1 is substituted by O atom). Other calculations show either formation of N vacancies in positions (1)-(3) or N atom substitution by O atom in positions (2) and (3).

## 4 Results and discussion

For three considered positions (1 to 3 in Fig.1) we estimated the N vacancy formation energy as well as oxygen incorporation and solution energy, respectively.

The formation energy $E_{form}^{N\_vac}$ of N vacancy was calculated as

$$E_{form}^{N\_vac} = E_{GB(N\_vac)} + \mu_{N\_atom} - E_{GB}, \qquad (1),$$

the oxygen incorporation energy $E_{inc}$ for the first time was suggested by Grimes and Catlow [12]:

$$E_{inc} = E_{GB(O\_inc)} - E_{GB(N\_vac)} - \mu_{O\_atom}, \qquad (2),$$

and oxygen solution energy $E_{sol}$ was calculated as

$$E_{sol} = E_{inc} + E_{form}^{N\_vac}, \qquad (3)$$

where $E_{GB(N\_vac)}$ is the total energy of a fully relaxed grain boundary containing one N vacancy in positions (1), (2) or (3) as depicted in Fig. 1, $E_{GB(O\_inc)}$ is the total energy for the grain boundary with O atom incorporated into the N vacancy, $E_{GB}$ is the same for grain boundary without defect, and, finally, $\mu_{N\_atom}$ ($\mu_{O\_atom}$) is the energy of an atom in the $N_2$ ($O_2$) molecule.

*Table 1.* Formation energies $E_{form}^{N\_vac}$ (in eV) for N vacancy in the grain boundary for three different positions (Fig. 1) and on the UN (001) and (110) surface (outermost) and central layers and as well as the average spin magnetic moment $\mu_{av}$ of U atoms in these systems.

| positions in Fig. 1. | Grain boundary | | Number of layers in slab | Surface layer | | | | Central layer | | | |
|---|---|---|---|---|---|---|---|---|---|---|---|
| | | | | (001) surface | | (110) surface | | (001) surface | | (110) surface | |
| | $E_{form}^{N\_vac}$ | $\mu_{av}(\mu_B)$ | | $E_{form}^{N\_vac}$ | $\mu_{av}(\mu_B)$ | $E_{form}^{N\_vac}$ | $\mu_{av}(\mu_B)$ | $E_{form}^{N\_vac}$ | $\mu_{av}(\mu_B)$ | $E_{form}^{N\_vac}$ | $\mu_{av}(\mu_B)$ |
| (1) | 3.47 | 1.46 | 7, 2×2 | 3.70 | 1.55 | 3.03 | 1.59 | 4.43 | 1.49 | 4.52 | 1.51 |
| (2) | 3.34 | 1.46 | 9, 2×2 | 3.70 | 1.45 | 3.04 | 1.51 | 4.42 | 1.41 | 4.35 | 1.49 |
| (3) | 3.48 | 1.46 | 11, 2×2 | 3.71 | 1.39 | 3.03 | 1.45 | 4.42 | 1.36 | 4.42 | 1.43 |
| | | | 7, 3×3 | 3.65 | 1.49 | 2.97 | 1.50 | 4.42 | 1.49 | 4.56 | 1.47 |

The N vacancy formation energies $E_{form}^{N\_vac}$ are summarized in Table 1. The results for the N vacancies in grain boundaries are compared with previous results for the N vacancies on the UN (001) and (110) surfaces [4]. We clearly see spin-changes similar for both UN surface and grain boundary. The formation energies $E_{form}^{N\_vac}$ in the GB are 3.3-3.5 eV. These values are comparable with analogous values for UN (001) (3.6-3.7 eV) and (110) (~3.0 eV) surfaces but are smaller than those in the bulk material (~4.4 eV) or in the (001) or (110) slabs central layer (4.3-4.6 eV). It indicates a clear trend for segregation of vacancies towards the grain boundaries and surfaces.

*Table 2.* Incorporation energy $E_{inc}$ and solution energy $E_{sol}$ (in eV), average spin magnetic moment of U atom $\mu_{av}$ as well as effective charge of incorporated O atoms into the UN (001) and (110) surface (outermost) atomic layer, central layer of UN(001)slab as well as grain boundary (Figure 1).

| | O incorporation into grain boundary | | | | O incorporation into central layer of (001) slab | | | |
|---|---|---|---|---|---|---|---|---|
| positions in Fig. 1. | $E_{inc}$ | $E_{sol}$ | $\mu_{av}(\mu_B)$ | $q_{eff}(e^-)$ | Number of layers | $E_{inc}$ | $E_{sol}$ | $\mu_{av}(\mu_B)$ | $q_{eff}(e^-)$ |
| (1) | -5.92 | -2.45 | 1.45 | -1.41 | 7, 2×2 | -6.61 | -2.18 | 1.47 | -1.42 |
| (2) | -5.67 | -2.33 | 1.44 | -1.38 | 9, 2×2 | -6.61 | -2.19 | 1.39 | -1.38 |
| (3) | -5.92 | -2.44 | 1.44 | -1.36 | 7, 3×3 | -6.60 | -2.18 | 1.45 | -1.42 |

| | O incorporation into surface layer of slab | | | | | | | |
|---|---|---|---|---|---|---|---|---|
| | (001) surface | | | | (110) surface | | | |
| Number of layers | $E_{inc}$ | $E_{sol}$ | $\mu_{av}(\mu_B)$ | $q_{eff}(e^-)$ | $E_{inc}$ | $E_{sol}$ | $\mu_{av}(\mu_B)$ | $q_{eff}(e^-)$ |
| 7, 2×2 | -6.18 | -2.48 | 1.495 | -1.36 | -5.82 | -2.79 | 1.52 | -1.29 |
| 9, 2×2 | -6.19 | -2.48 | 1.41 | -1.36 | -5.82 | -2.78 | 1.47 | -1.29 |
| 11, 2×2 | -6.195 | -2.48 | 1.365 | -1.35 | -5.82 | -2.79 | 1.42 | -1.29 |
| 7, 3×3 | -6.13 | -2.48 | 1.46 | -1.36 | -5.75 | -2.78 | 1.47 | -1.28 |

Oxygen incorporation energy $E_{inc}$ and solution energy $E_{sol}$ are summarized in Table 2. These energies for the vacancies in the grain boundaries (-5.6÷-5.9 vs. -2.3÷-2.5 eV, respectively) are close to those for the UN surface (-5.7÷-6.6 eV for O atom incorporation and -2.4÷-2.8 for O atom solution).

## 5 Conclusion

*Ab initio* DFT method was used to analyze the role of point defects and O interaction with UN grain boundaries using (310)[001](36.8°) tilt model. We obtained for the first time the energies of nitrogen vacancy formation and O atom incorporation into N vacancy in the GB and compared these results with previously obtained results for both UN(001) and (110) surfaces. We have observed a clear evidence for vacancy segregation at the interfaces, open surfaces and grain boundaries. Overall, defects energetics of the UN grain boundary and surface very well correlate with each other.

**Acknowledgements:** The authors thank A.L. Shluger and K. McKenna for help in constructing the grain boundary model, as well as A. Kuzmin, Yu.A. Mastrikov and S. Piskunov for a technical assistance. D.B. is grateful to European Social Fund project No. 2009/0216/1DP/1.1.1.2.0/09/APIA/VIAA/044.